\documentclass{article}

\usepackage{arxiv}

\usepackage[utf8]{inputenc} 
\usepackage[T1]{fontenc}    
\usepackage{hyperref}       
\usepackage{url}            
\usepackage{booktabs}       
\usepackage{amsfonts}       
\usepackage{nicefrac}       
\usepackage{microtype}      
\usepackage{lipsum}
\usepackage{graphicx}
\graphicspath{ {./images/} }

\title{Noise-tolerant wavefront shaping in a Hadamard basis}

\author{
 Bahareh Mastiani \\
  Biomedical Photonic Imaging Group\\
  Faculty of Science and Technology\\
  University of Twente \\
  P.O. Box 217, 7500 AE Enschede, The Netherlands\\
  \texttt{b.mastiani@utwente.nl} \\
   \And
 Ivo M. Vellekoop \\
  Biomedical Photonic Imaging Group\\
  Faculty of Science and Technology\\
  University of Twente \\
  P.O. Box 217, 7500 AE Enschede, The Netherlands\\
  \texttt{i.m.vellekoop@utwente.nl} \\
 
}
\usepackage{amsmath}  
\usepackage{xspace}
\begin{document}
\maketitle
\newcommand{\mum}{\text{\textmu m}}
\newcommand{\abs}[1]{\left\lvert#1\right\rvert}
\newcommand{\Kubby}{Tao\xspace}
\newcommand{\Gigan}{Popoff\xspace}
\begin{abstract}
Light scattering is the main limitation for optical imaging. However, light can be focused through or inside turbid media by spatially shaping the incident wavefront.
Wavefront shaping is ultimately limited by the available photon budget. We developed a new `dual reference' wavefront shaping algorithm that optimally uses the available light. Our method allows for multi-target wavefront shaping, making it suitable for transmission matrix measurements or transmitting images. We experimentally confirmed the improvement of the focus intensity compared to existing methods.
\end{abstract}


\section{Introduction}
Scattering and diffusion of light are the main limitations for optical imaging. However, wavefront shaping techniques make it possible to shape the incident wavefront so that it exactly matches the scattering properties of the medium, and form a focus through or inside it \cite{kubby2019wavefront}. Feedback-based wavefront shaping algorithms search for the optimum incident field that maximizes some feedback signal, such as transmission into a specific output mode \cite{vellekoop2015feedback}. The task of finding the optimum wavefronts for multiple targets (output modes) is equivalent to measuring the transmission matrix (TM) of the scattering medium, up to an unknown phase factor for each row of the matrix\cite{popoff2010image}. The TM describes the input-output response of the scattering medium. Knowledge of this matrix makes it possible to focus at any arbitrary position behind the medium \cite{yu2013measuring, kim2015transmission}, to transmit images through it \cite{popoff2010image, mosk2012controlling}, or to study fundamental aspects of light scattering, such as the distribution of transmission channels \cite{yilmaz2019transverse, rotter2017light}.

\begin{figure}
    \centering
    \includegraphics{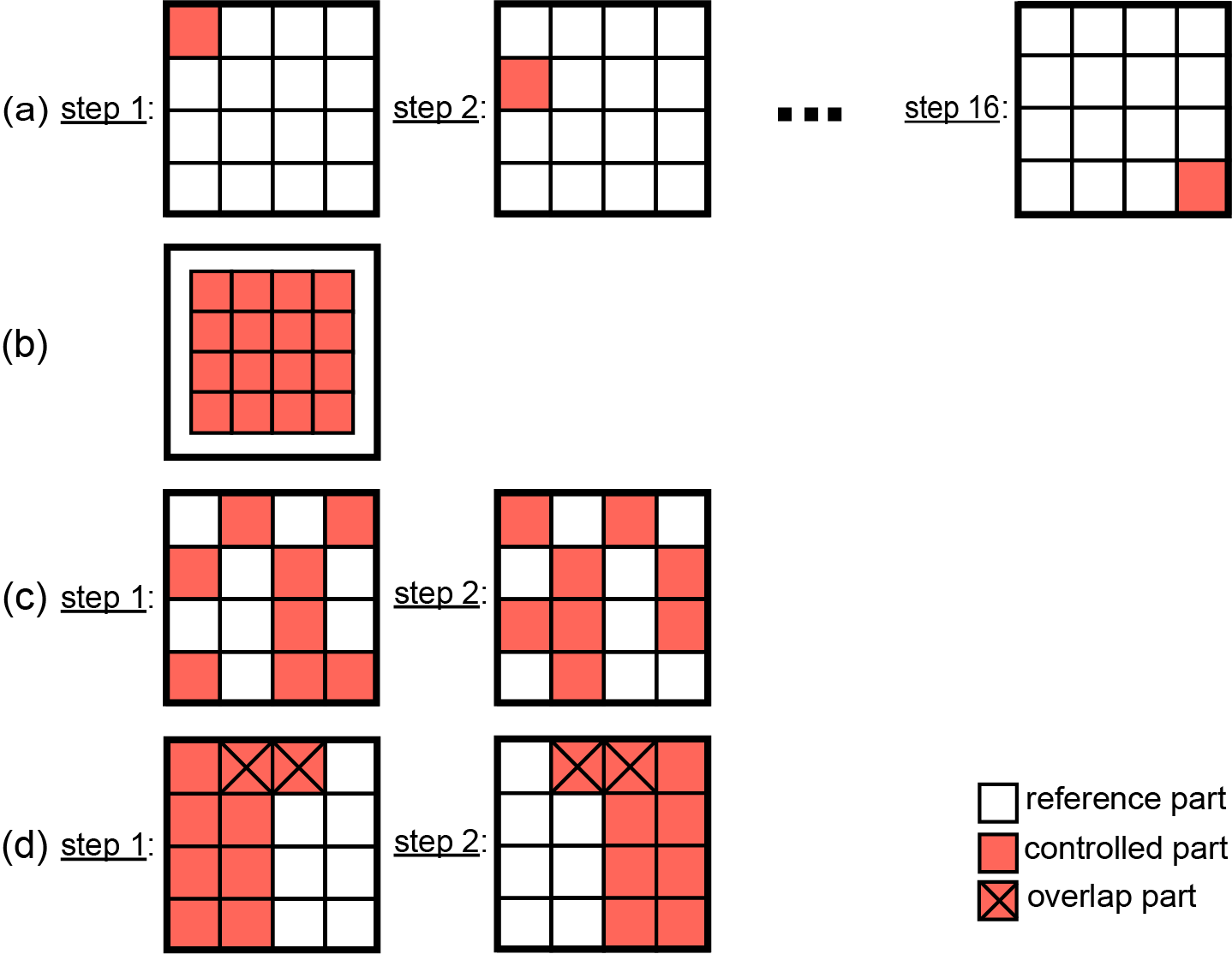}
    \caption{Feedback based wavefront shaping algorithms. (a) Stepwise sequential algorithm: a single segment (in red) is controlled in every step. (b) \Gigan et al. [2010] algorithm: always 65\% of the SLM segments are controlled. (c) \Kubby et al. [2017] algorithm: in every step half of the segments are controlled. (d) Dual reference algorithm: First, a group of segments is modulated (in red), and in the second step the rest of the segments plus some overlap segments (crossed) are modulated.}
    \label{fig:algorithms}
\end{figure}

A common procedure to find the optimum incident field with a phase-only spatial light modulator (SLM) is using the stepwise sequential algorithm (SSA) \cite{vellekoop2015feedback}, as illustrated in Fig. \ref{fig:algorithms}(a). In this algorithm, the SLM is divided into segments, and the phase of each individual segment is varied between 0 and 2$\pi$ consecutively while keeping the phase of all other segments fixed. Due to interference between light originating from the controlled segment and light originating from all other segments, the feedback signal will respond sinusoidally. By fitting these sinusoids, the TM is reconstructed \cite{vellekoop2015feedback}.

The drawback of the SSA algorithm becomes apparent when many segments are used. In this case the contribution of each segment is small compared to the reference field coming from the rest of the segments, giving a low interferometric visibility of the feedback signal. This results in a low signal to noise ratio (SNR) in the measured TM elements.
This low SNR is especially noticeable when the photon budget for the measurements is limited, such as in high-speed wavefront shaping \cite{feldkhun2019focusing}, and in microscopy. 

A method that was introduced by \Gigan et al. \cite{popoff2010measuring} organizes the segments in a Hadamard basis. The feedback signal is measured for each Hadamard vector, allowing the TM to be reconstructed. In this method, always a large fraction of the segments is modulated, increasing the visibility of the interference signal, and consequently improving the SNR \cite{popoff2010measuring}. However, this technique requires part of the incident light to remain unmodulated serving as a reference field for the measurements. Since part of the light is not modulated, the contrast of the final focus is not optimal. This method is illustrated in Fig. \ref{fig:algorithms}(b), showing 65\% of the SLM are always modulated (in red) and the other 35\% of the segments are the reference part \cite{popoff2010measuring}.

A different way to increase the SNR for a given photon budget is by performing a pre-optimization \cite{yilmaz2013optimal, tao2017three}. In pre-optimization schemes, one first performs an optimization with part of the SLM segments and uses that solution as a starting point for a second optimization step. With pre-optimization, the feedback signal in the second optimization step will be higher, causing the second step to be more robust to noise in most cases \cite{yilmaz2013optimal}. 

A special case of pre-optimization was developed by \Kubby et al.\cite{tao2017three}. In the first step, half of the segments is modulated using a Hadamard-based algorithm while the other half is used as a reference. Since the number of controlled segments is equal to the number of reference segments, this approach leads to an optimally balanced interference of modulated and reference field. In the second step, the optimized half of the wavefront is displayed on the SLM. The roles of modulated and reference segment are now switched (see Fig. \ref{fig:algorithms}(c)), and the other half of SLM segments is optimized. 

The method of \Gigan et al. can be used to simultaneously find the wavefronts to focus light to multiple targets, or even to project arbitrary images through the scattering medium. In effect, this method measures multiple rows of the TM simultaneously. Unfortunately, pre-optimization approaches can not be used for this purpose. After the pre-optimization step, one has to choose what wavefront to use as a starting point for the next step, so the procedure can only be used on one point at a time.

In this work, we present a new algorithm, called dual reference, which is a combination of the methods used by \Gigan et al. \cite{popoff2010image}, and \Kubby et al. \cite{tao2017three}. We take advantage of the optimal balanced interference of the modulated and reference field as in \Kubby et al., while keeping the ability to measure all rows of the TM simultaneously, as in \Gigan et al.

We will first describe our dual reference method, and explain why it outperforms other methods for a wide range of photon budgets. Afterward, we compare the experimental results of the performance of the proposed method with the existing feedback-based wavefront shaping methods for various photon budgets.

\section{Dual reference algorithm}
In our dual reference algorithm, we split the SLM segments into two groups of the same size with a small number $O$ of overlapping segments, as illustrated in Fig.~\ref{fig:algorithms}(d). The groups have size $N_1=(N+O)/2$, where $N_1$ is a power of two. The field in the $b$-th output mode is now given by 

\begin{equation}
E_b = \sum_{a=1}^N t_{ba}E_a = \sum_{a=1}^{N_1} t_{ba}E_a + \sum_{a=N_1+1}^N t_{ba}E_a,
\label{eq:outputfield}
\end{equation}
where $E_a$ is the field in the $a$-th input mode, $N$ is the total number of SLM segments, and $t_{ba}$ are the elements of the transmission matrix. In the first step, the algorithm of \Gigan et al. \cite{popoff2010measuring} is used: a Hadamard pattern is displayed on the segments of group 1 (segments $1$ to $N_1$) and the remaining segments are used as a reference. The relative phase between the two groups is changed from 0 to $2\pi$ in $P$ steps. The measured feedback signal for Hadamard vector $j\in [1,N_1]$ and phase step $k\in [1,P]$ is given by 

\begin{equation}
I_{jk} = \abs{e^{i\phi_k}\sum_{a=1}^{N_1}
t_{ba} h_{ja} E_0 + E_\text{ref,2}}^2.
\label{eq:intensity}
\end{equation}
where $h_{ja}\in\{-1,1\}$ is the $a$-th element of the (non-normalized) $j$-th Hadamard basis vector, $E_0$ is a constant to normalize the incident field, and the reference field from group 2 is given by $E_\text{ref,2}\equiv \sum_{a=N_1+1}^N t_{ba}E_a$. We omitted the subscript $b$ for brevity. Substituting $E_j \equiv \sum_{a=1}^{N_1}t_{ba}h_{ja}E_a$ gives

\begin{equation}
I_{jk} = \abs{E_\text{ref,2}}^2+\abs{E_j}^2+E_\text{ref,2}^* E_j e^ {i\phi_k} +  E_\text{ref,2} E_j^* e^ {-i\phi_k}.
\label{eq:expanded intensity}
\end{equation}

In order to retrieve the optimized wavefront, we first perform a Fourier transform of the measured intensity
\begin{equation}
S_j = \frac{1}{P}\sum_{k=1}^P I_{jk} e^{-i\phi_k} = E_\text{ref,2}^* E_j,
\label{eq:sum intensity}
\end{equation}
Finally, a Hadamard transform is used to convert $S_j$ to $S_{a1}= E_\text{ref,2}^* t_{ba} E_{a}$. Assuming that we have an even illumination on the SLM, this gives us the transmission matrix elements $t_{ba}$ for the elements of group $1$ (elements $1$ to $N_1$), up to an unknown overall factor $E_\text{ref,2}^*$.

Our modification to \Gigan`s algorithm is to add a second step, where group 1 is now used as a reference, and the segments of group 2 are modulated. These two steps are shown in Fig. \ref{fig:algorithms} (d) with the modulated SLM segments in red and the overlap segments as crossed segments. 
Since the number of controlled segments is close to the number of reference segments for all measurements, the modulated and reference fields have almost equal average contribution to the field at point $b$. This 50/50 choice for the reference and modulated part guarantees the maximal interferometric visibility and minimal sensitivity to noise for a given photon count.

After completing the two steps, we have the corrected wavefronts for group 1 and group 2. To find the final optimal wavefront, these two partial corrections need to be combined. However, since the two groups used different reference fields in the measurements, the two partial wavefronts will have a different phase offset. Since the overlap segments are present in both groups, these segments can be used for determining the relative phase between the two corrections. Denoting the overlap segments as $a=1..O$, we compute the relative phase from the dot product of the two measurements for these overlapping segments
\begin{equation}
    \arg \left(E_\text{ref,2}\right) - \arg \left(E_\text{ref,1}\right) = \arg \sum_{a=1}^O S_{a1} S_{a2}^*
\end{equation}
and the relative amplitude follows from
\begin{equation}
    \frac{\abs{E_\text{ref,2}}^2}{\abs{E_\text{ref,1}}^2}=\frac{\sum_{a=1}^O \abs{S_{a1}}^2}{\sum_{a=1}^O \abs{S_{a2}}^2}
\end{equation}
After compensating the relative phase and amplitude of the two partial wavefronts, these two wavefronts can be combined into a single wavefront.

\section{Experiment}
\begin{figure}
\centering
\includegraphics[width=\linewidth]{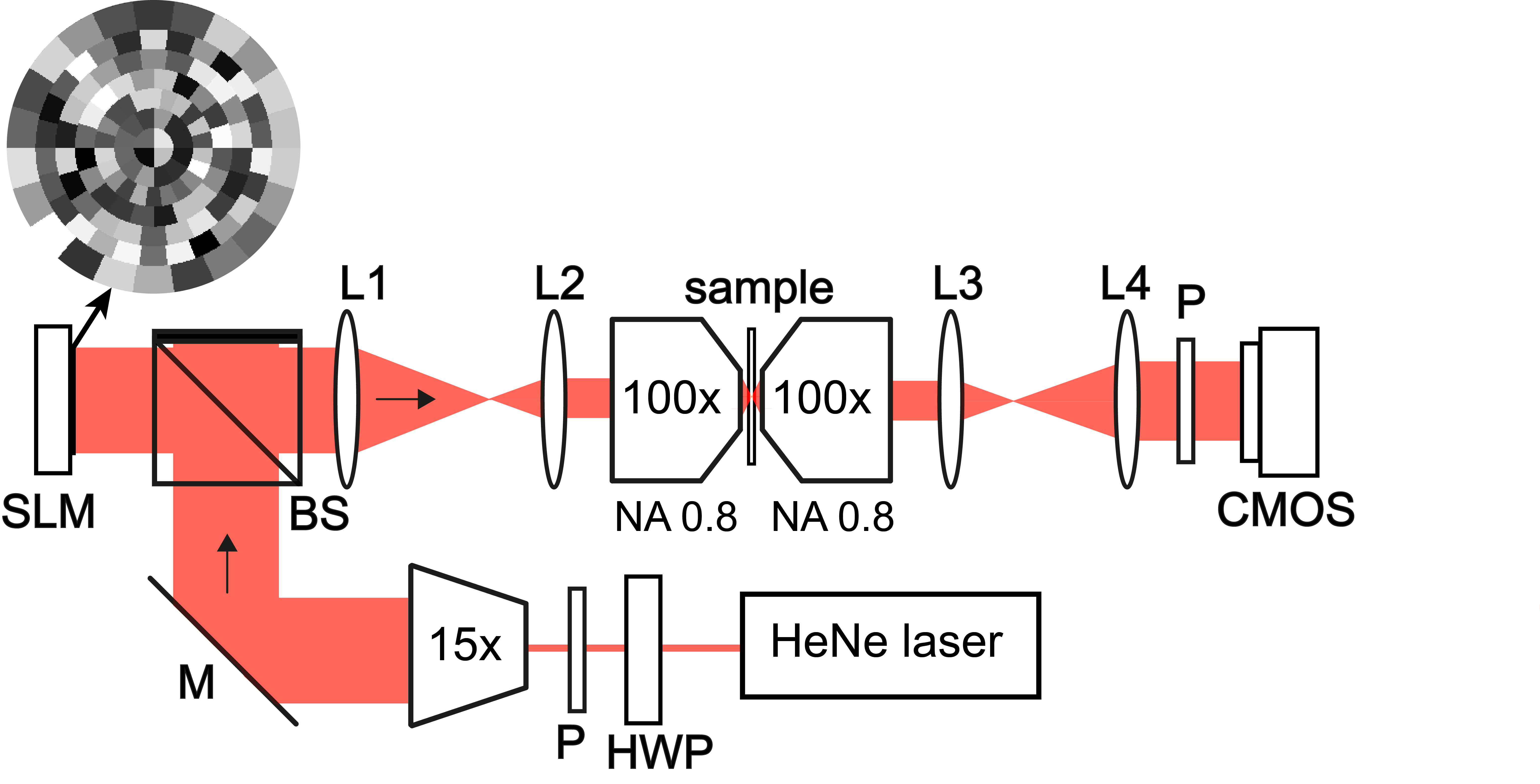}
\caption{Schematic of the experimental setup. HWP, half-wave plate; M, mirror; BS, 50\% non-polarizing beam splitter; P, polarizer; CMOS, complementary metal oxide semiconductor camera; L1, L2, L3 and L4, lenses with focal length of respectively 200 mm, 75 mm, 75 mm and 150 mm. The inset shows a random wavefront displayed on the modulating segments of the SLM.}
\label{fig:setup}
\end{figure}

Now, we will experimentally compare the performance of the dual reference algorithm to the original SSA algorithm and to the methods of \Gigan and \Kubby. Since \Kubby's method can only measure one row of the TM at the time (hence only focus on one target), the comparison with that method is not completely fair. For completeness, we do include the method in the comparison, using only a single target in the measurements.

The experimental setup is illustrated in Fig. \ref{fig:setup}. In this setup, light from a HeNe laser (Thorlabs, 2mW, $\lambda$ = 632.8 nm) is expanded and modulated by a phase-only spatial light modulator (Hamamatsu X13138-07). A 4f system images the SLM onto the back focal plane of a microscope objective (A-Plan 100x/0.8, Zeiss), which focuses the light onto the sample. After transmitting through the sample, the light is collected by an identical objective lens and recorded by a CMOS camera (Basler acA640-750um). The camera records the intensity distribution at the back focal plane of the second objective. A set of 100 independent targets on the camera is used as the feedback for the wavefront shaping algorithms. 

The sample is an 11$\pm$3 $\mum$ thick layer of zinc-oxide (Sigma Aldrich, average grain size 200 nm) on a coverslip with a thickness of 170 $\mum$. The transport mean free path of similar zinc oxide samples was measured to be around 0.6 $\mum$ at a wavelength of $\lambda$ = 632.8 nm \cite{vanPutten}. Consequently, the sample is optically thick so that there is no transmitted ballistic light. The sample is mounted on a translation stage (Zaber T-LSM050A) to scan the sample laterally for ensemble averaging. 

The beam illuminating the SLM has a Gaussian intensity profile. It is well known that such a non-uniform illumination causes a sub-optimal enhancement\cite{akbulut2011focusing}. To overcome this effect, we choose a circular geometry instead of the common rectangular one, as is shown in Fig. \ref{fig:setup}(inset). The sizes of the rings are adjusted such that each segment, on average, contributes equally to the target signal. This way, the illumination effectively is uniform, and the enhancement is maximized.

To control the photon budget used in the experiments, we used various neutral density filters with different transmission ratios in front of the camera. The power of detected light was measured by a calibrated power meter (S121C, Thorlabs). The measured power is converted to the photon count collected by the camera. We kept increasing the total photon budget as long as the camera did not reach saturation during the measurement.

In the wavefront shaping experiments, we measured the enhancement of the focus, defined as the ratio of the optimized intensity at the focus location to the reference intensity which is the averaged intensity over 100 positions for the sample \cite{kubby2019wavefront}.

In our experiments, measuring the corrected wavefront for one target took 50 seconds. For more than one target, this time interval does not change for all the algorithms except for \Kubby's method, which is linearly increased with the number of targets. 

\section{Optimal number of phase steps}
All four algorithms require the choice of two parameters: the number of SLM segments ($N$) and the number of phase steps per segment ($P$). Together, these parameters determine the total number of measurements $NP$. It is well known that in absence of noise the expected enhancement scales linearly with $N$ \cite{vellekoop2007focusing}, so $N$ should be chosen as high as possible, resulting in a minimal number of phase steps $P=3$. In presence of noise, however, the relation is more complicated, and different types of noise (such as read-out noise, shot noise and laser excess noise) affect the results in different ways\cite{yilmaz2013optimal}. When the number of measurements $NP$ is fixed, a higher value of $P$ will give a more accurate measurement of each transmission matrix element, at the expense of decreasing $N$. 

To find the optimum value for $P$ for the different algorithms, we measured the enhancement for different combinations of $P$ and $N$. To enable a fair comparison, we choose the number of SLM segments in a way that all algorithms have the same number of measurements. For the lowest phase steps number ($P=$3), we set 128 segments for \Kubby's method \cite{tao2017three} and SSA. For \Gigan's method \cite{popoff2010image} the SLM is divided into 197 segments, out of which 128 segments are modulated, and the remaining 69 segments (35\%) serves as the reference. For our dual reference algorithm, we had 123 modulated segments with 5 segments mutual in two 64-segment groups. As a result, the number of measurements for all the algorithms is the same, and equal to 384. We kept this number fixed by reducing the number of segments by two or four when $P$ is respectively 6 and 12. 

\begin{figure}
\centering\includegraphics[width=\linewidth]{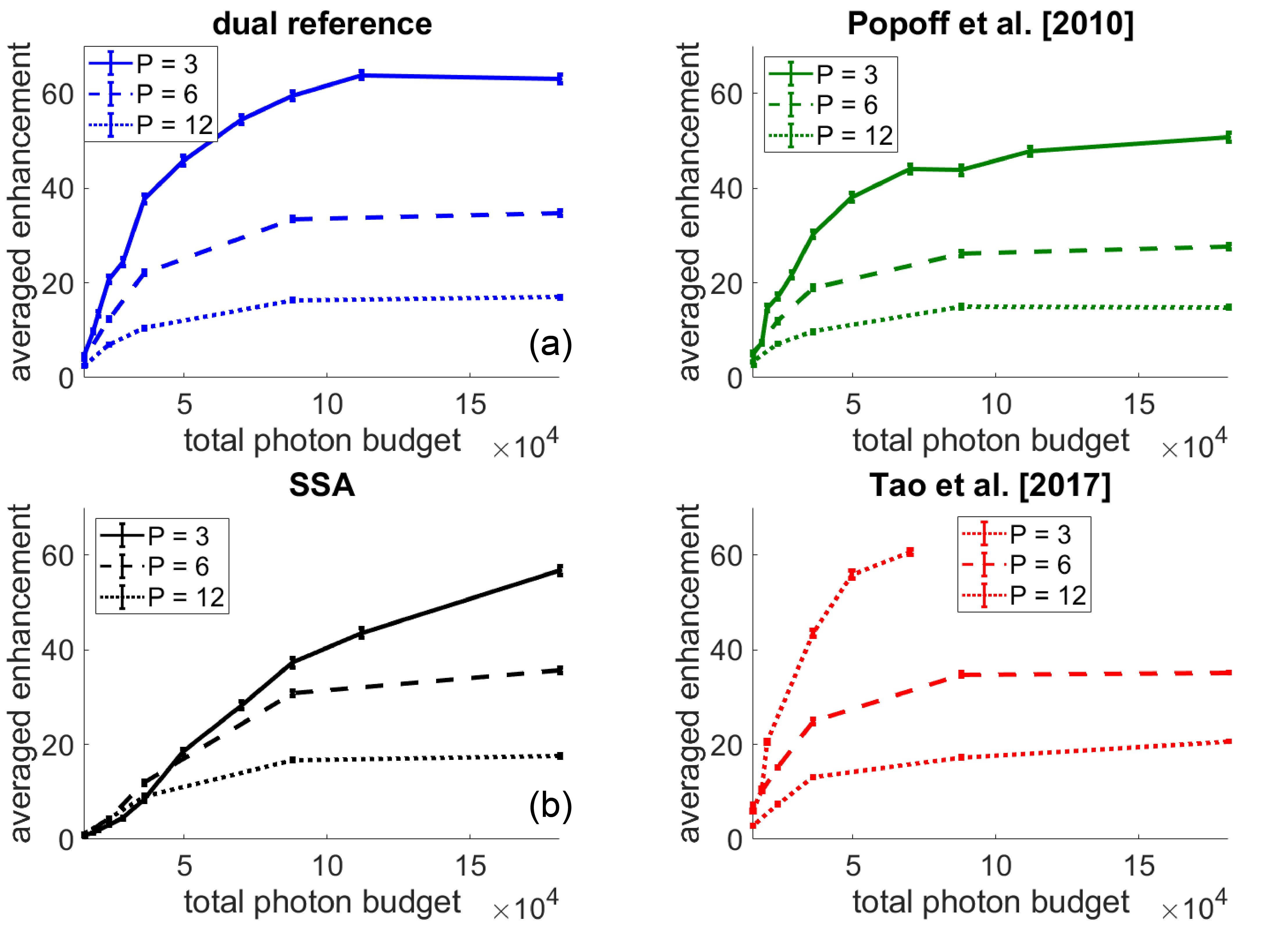}
\caption{The measured averaged enhancement versus the total photon budget when the number of phase steps is 3 (solid), 6 (dash), and 12 (dot), (a) for dual reference algorithm, (b) SSA, (c) \Gigan et al. \cite{popoff2010measuring}, and (d) \Kubby et al. \cite{tao2017three}. Bars represent the standard error of the measurement set.}
\label{fig:result1}
\end{figure}

 Figure \ref{fig:result1} shows the averaged enhancement over 100 targets and 3 trials versus total photon budget when $P$ is 3 (solid curve), 6 (dash curve), and 12 (dot curve). For the dual reference method Fig. \ref{fig:result1}(a), \Gigan's method Fig. \ref{fig:result2}(b), and \Kubby's method Fig. \ref{fig:result2}(d), the highest enhancement for every photon budget is obtained for $P=3$. However, Fig. \ref{fig:result2}(c) shows that for a photon budget lower than $5\cdot 10^4$, SSA shows a slightly better performance for larger values of $P$.

The results from Fig. \ref{fig:result1} show that a lower $P$ with higher $N$ is better than performing accurate measurements for fewer segments. Therefore, we set $P=3$ for all algorithms and compare the performance of the dual reference algorithm with the other algorithms.

\section{Results and discussion}
\begin{figure}
\centering
\includegraphics[width=\linewidth]{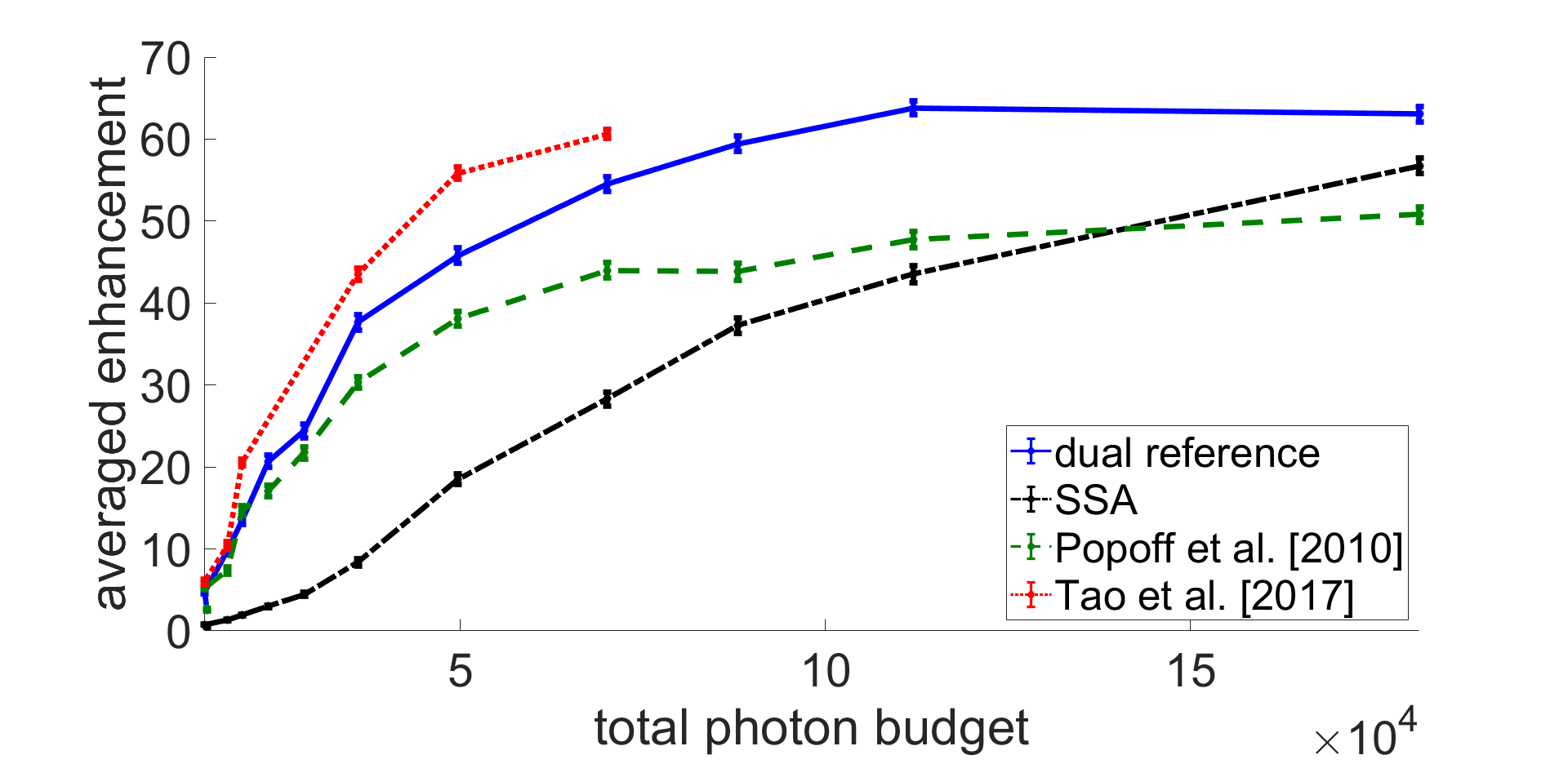}
\caption{The measured averaged enhancement versus the total photon budget for dual reference algorithm (blue solid curve), SSA (black dash-dot curve), \Gigan et al. \cite{popoff2010measuring} (green dash curve), and \Kubby et al. \cite{tao2017three} (red dot curve). Bars represent the standard error of the measurement set.  }
\label{fig:result2}
\end{figure}

Figure \ref{fig:result2} illustrates the measured enhancement for the different methods, averaged over 100 targets and 3 trials. For all algorithms, the enhancement increases as the total photon budget increases since the signal to noise ratio increases.     

The measured enhancement for \Kubby's method, shown in the red dot curve in Fig. \ref{fig:result2}, is higher than the other algorithms due to the pre-optimization step. However, finding the corrected wavefronts for all 100 targets took 100 times longer than the other three algorithms. This is because the corrected wavefront from step 1 is different for every target and it prevents us from simultaneous multi-target optimization. During step 2, the corrected wavefront from step 1 is displayed on the SLM, causing the camera to be overexposed for higher photon budgets. For this reason, there are fewer reported measurements for \Kubby's method than for the other methods.

The measured enhancement for the dual reference algorithm, shown as the blue solid curve in Fig. \ref{fig:result2}, confirms that using this method gives us higher enhancement for multi-target optimization compared to SSA (black dash-dot curve) and \Gigan's method (green dash curve). 

For a high photon budget, we theoretically expect to have an enhancement of $N\pi/4$ for SSA, dual reference, and \Kubby's methods, and $0.65 N\pi/4$ for \Gigan's method. In our experiments, the measured enhancement for the highest photon budget ($1.8\cdot 10^5$ photons) is 36 \% lower than the expected enhancement from the theory. This difference can be explained by practical imperfections, such as the static (non-modulated) field caused by the reflection of the SLM front surface. In our case, this static field has an amplitude of 10 \% of the input field amplitude.

\section{Conclusion}
With the advance of high-speed SLMs, the photon budget will ultimately be the limiting factor for wavefront shaping. We have presented a new feedback-based wavefront shaping algorithm to optimally use the available light.

We demonstrated that for a fixed number of measurements the best performance of all algorithms is reached by minimizing the number of phase steps.

Our algorithm achieves the maximal interferometric visibility during the measurements, resulting in an optimal SNR. Moreover, there is no need to reserve segments for the reference beam. We have experimentally demonstrated that this method enables us to perform simultaneous multi-target optimization with a higher enhancement than the popular methods used by \Gigan et al. [2010] and \Kubby et al. [2017] (for multi-target) for a wide range of photon budgets. 

We envision that the presented method could be beneficial to the experiments having the transmission matrix measurement involved, especially when the photon budget is limited.  

\section*{Funding}
This work was financially supported by the Nederlandse Organisatie voor Wetenschappelijk Onderzoek (TTW-NWO, Vidi grant 14879).

\section*{Disclosures}
The authors declare no conflicts of interest.

\bibliographystyle{unsrt}  
\bibliography{references} 

\end{document}